\documentclass[
aps,
prl,
reprint,
superscriptaddress,
amsfonts,
amssymb,
amsmath,
showpacs,
eprint
]{revtex4-1}

\usepackage{graphicx}
\usepackage{dcolumn}
\usepackage{bm}
\usepackage{ulem}
\usepackage{framed}
\usepackage{hyperref}
\usepackage{color}
\usepackage{braket}

\usepackage{sidecap}
\usepackage{float}

\def\sm{\sigma^-}
\def\smdag{\sigma^+}

\def\r1{\textbf{r}}

\def\a{{a}}
\def\adag{{a}^\dagger}
\newcommand{\abs}[1]{\left| #1 \right|} 

\usepackage{makeidx}
\makeindex

\begin{document}

\abovedisplayskip=8pt
\abovedisplayshortskip=8pt
\belowdisplayskip=6pt
\belowdisplayshortskip=6pt


\title{
Cavity-enhanced simultaneous dressing of quantum dot exciton and biexciton states
}

\author{F.~Hargart}
 \email{f.hargart@ihfg.uni-stuttgart.de}
\affiliation{Institut f\"ur Halbleiteroptik und Funktionelle Grenzfl\"achen, Research Center SCoPE and IQST, Universit\"at Stuttgart, Allmandring 3, 70569 Stuttgart, Germany}

\author{M.~M\"uller}
\affiliation{Institut f\"ur Halbleiteroptik und Funktionelle Grenzfl\"achen, Research Center SCoPE and IQST, Universit\"at Stuttgart, Allmandring 3, 70569 Stuttgart, Germany}

\author{K.~Roy-Choudhury}
 \email{kroy@physics.queensu.ca}
\affiliation{Department of Physics, Engineering Physics and Astronomy, Queen's University, Kingston, Ontario, Canada K7L 3N6}

\author{S.~L.~Portalupi}
\affiliation{Institut f\"ur Halbleiteroptik und Funktionelle Grenzfl\"achen, Research Center SCoPE and IQST, Universit\"at Stuttgart, Allmandring 3, 70569 Stuttgart, Germany}

\author{C.~Schneider}
\affiliation{Technische Physik and Wilhelm Conrad R\"ontgen Research Center for Complex Material Systems, Physikalisches Institut, Universit\"at W\"urzburg, Am Hubland, 97074 W\"urzburg, Germany}

\author{S.~H\"ofling}
\affiliation{Technische Physik and Wilhelm Conrad R\"ontgen Research Center for Complex Material Systems, Physikalisches Institut, Universit\"at W\"urzburg, Am Hubland, 97074 W\"urzburg, Germany}
\affiliation{SUPA, School of Physics and Astronomy, University of St. Andrews, St. Andrews, KY16 9SS, United Kingdom}

\author{M.~Kamp}
\affiliation{Technische Physik and Wilhelm Conrad R\"ontgen Research Center for Complex Material Systems, Physikalisches Institut, Universit\"at W\"urzburg, Am Hubland, 97074 W\"urzburg, Germany}

\author{S.~Hughes}
\affiliation{Department of Physics, Engineering Physics and Astronomy, Queen's University, Kingston, Ontario, Canada K7L 3N6}

\author{P.~Michler}
\affiliation{Institut f\"ur Halbleiteroptik und Funktionelle Grenzfl\"achen, Research Center SCoPE and IQST, Universit\"at Stuttgart, Allmandring 3, 70569 Stuttgart, Germany}

\date{\today}

\begin{abstract}
We demonstrate the simultaneous dressing of both vacuum-to-exciton and exciton-to-biexciton transitions of a single semiconductor quantum dot in a high-Q micropillar cavity, using photoluminescence spectroscopy. Resonant two-photon excitation of the biexciton is achieved by spectrally tuning the quantum dot emission with respect to the cavity mode. The cavity couples to both transitions and amplifies the Rabi-frequency of the likewise resonant cw laser, driving the transitions. We observe strong-field splitting of the emission lines, which depend on the driving Rabi field amplitude and the cavity-laser detuning. A dressed state theory of a driven 4-level atom correctly predicts the distinct spectral transitions observed in the emission spectrum, and a detailed description of the emission spectra is further provided through a polaron master equation approach which accounts for cavity coupling and acoustic phonon interactions of the semiconductor medium.
\end{abstract}


\maketitle


Single-photon emission from a resonantly driven quantum emitter has attracted increasing attention both from fundamental science perspectives and as a potential quantum light source for applications in quantum information science. Applying a strong continuous wave (cw) laser on a two-level transition facilitates hybrid light-matter systems, typically described in a dressed-state picture \cite{Mollow_1969,Cohen-Tannoudji.Dupont-Roc.ea_1998}. The study of dressed states in semiconductor quantum dots (QDs) and their emission properties has become a vital research field in semiconductor quantum optics. In previous work on dressed exciton-biexciton states in QDs, either the ground state-exciton transition ($\left|G\right>$-$\left|X\right>$) or the exciton-biexciton transition ($\left|X\right>$-$\left|XX\right>$) were dressed by a strong coupling laser \cite{Xu.Sun.ea_2007,Jundt.Robledo.ea_2008,Gerardot.Brunner.ea_2009,Muller.Fang.ea_2008,Flagg.Muller.ea_2009,Vamivakas.Zhao.ea_2009,Ates.Ulrich.ea_2009,He.He.ea_2013}. The dressed states were studied with two-color experiments, where a second weak laser probes the dressed states \cite{Xu.Sun.ea_2007,Jundt.Robledo.ea_2008,Gerardot.Brunner.ea_2009,Muller.Fang.ea_2008} or with resonance fluorescence studies, i.e., observing the characteristic Mollow triplet \cite{Flagg.Muller.ea_2009,Vamivakas.Zhao.ea_2009,Ates.Ulrich.ea_2009,He.He.ea_2013}.

Cascaded single-photon emission from the Mollow triplet sidebands of a QD \cite{Ulhaq.Weiler.ea_2012} and linewidth broadening \cite{Ulrich.Ates.ea_2011,Roy.Hughes_2011} due to excitation-induced dephasing have been demonstrated. Furthermore, the resonant coupling of a Mollow triplet sideband to an optical cavity has been studied \cite{Kim.Shen.ea_2014} and a decrease of the Mollow triplet sideband splitting was observed with increasing temperature, an effect, which was attributed to a phonon-induced renormalization of the driven dot Rabi frequency \cite{Wei.He.ea_2014}. In a recent study, Mollow quintuplets from coherently excited QDs have been presented \cite{Ge.Weiler.ea_2013}. Lately, a doubly-dressed single QD exposed to a bichromatic laser field has been studied thereby observing the nearly complete elimination of the resonance fluorescence spectral line at the driving laser frequency \cite{He.He.ea_2015}. Systematic studies of two-color second-order correlations of the light scattered near-resonantly by a QD have also been  presented \cite{Peiris.Petrak.ea_2015}.
In addition, two-color experiments where a second weak laser probes the dressed states generated by a strong laser field on the QD have been performed; e.g., a second probe laser was used to measure splitting of the dressed states by differential transmission \cite{Xu.Sun.ea_2007,Jundt.Robledo.ea_2008,Gerardot.Brunner.ea_2009} or for excitation of excess charge carriers in the barrier to generate QD photoluminescence (PL), thus revealing  dressed states \cite{Muller.Fang.ea_2008}.

In this work we demonstrate simultaneous dressing of ground state-exciton and exciton-biexciton transition in a semiconductor QD by tuning the frequency of a single drive laser close to half a biexciton energy $\left(\hbar\omega_{XX}/2\right)$. Such frequency tuning 
allows us to establish the two-photon resonance condition for direct, coherent excitation of the biexciton state. The optically prepared biexciton state $\ket{XX}$ can spontaneously decay via the $x$- or the $y$-polarized single exciton states ($\ket{X}$ or $\ket{Y}$). The current experiment uses an orthogonal excitation-measurement geometry, where the two $x$-polarized branches ($\left|XX\right>$-$\left|X\right>$-$\left|G\right>$) are coherently driven to prepare the biexciton state, and the $y$-polarized emission spectrum $S_{\rm y}$ is measured. Calculations using a simple 4-level atom model predicts 6, spectrally distinct transitions in the $y$-polarized spectrum $S_{\rm y}$, symmetric about the drive laser frequency which allows a direct observation of the dressed states created by the $x$-polarized drive laser. Strong enhancement of electric field of the driving laser within a high-Q micropillar cavity further aids the coherent dressing. A simple dressed-state model explains the qualitative features, and simulations using a more complete polaron master equation approach for the QD-cavity system shows excellent agreement with the experimental measurements.


\begin{figure}
        \centering
                \includegraphics[width=0.48\textwidth]{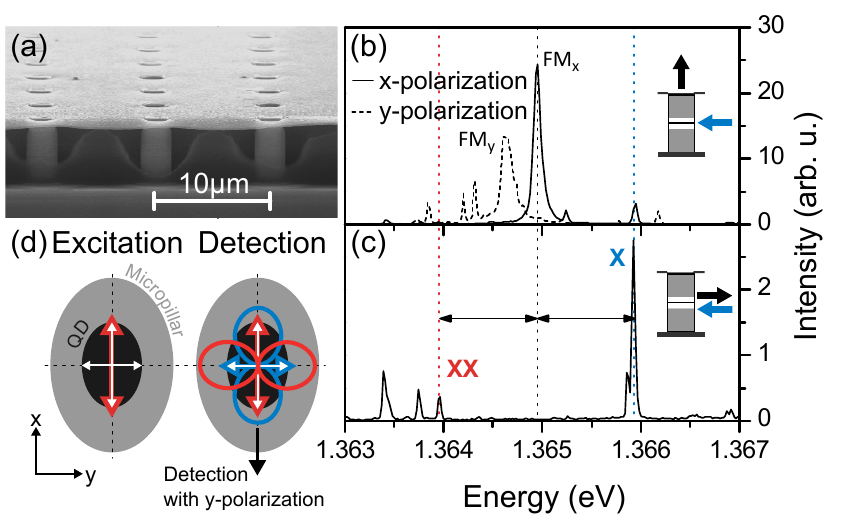}
        \caption{\label{Fig1}
        (Color online) (a) Scanning electron microscope image of the micropillar sample with an Al mask on top.
        (b) Polarization resolved spectra for top detection and above bandgap excitation.
        The x-polarized mode FM$_x$ is later used for cavity-enhanced resonant excitation. 
        (c) Side detection: The mode emission is suppressed whereas the biexciton XX and exciton X emission lines are clearly observable.
        (d) Excitation: The driving laser field is resonant to the x-polarized FM$_x$ and, thus, only drives the $\left|X\right>$ transitions of the QD. Detection: The resonantly excited biexciton state $\left|XX\right>$ can now decay via both channels $\left|X\right>$ and $\left|Y\right>$, but only the y-polarized emission including the exciton $\left|Y\right>$ state can be observed in our configuration.
        }
\end{figure}

\begin{figure}
        \centering
                \includegraphics[width=0.48\textwidth]{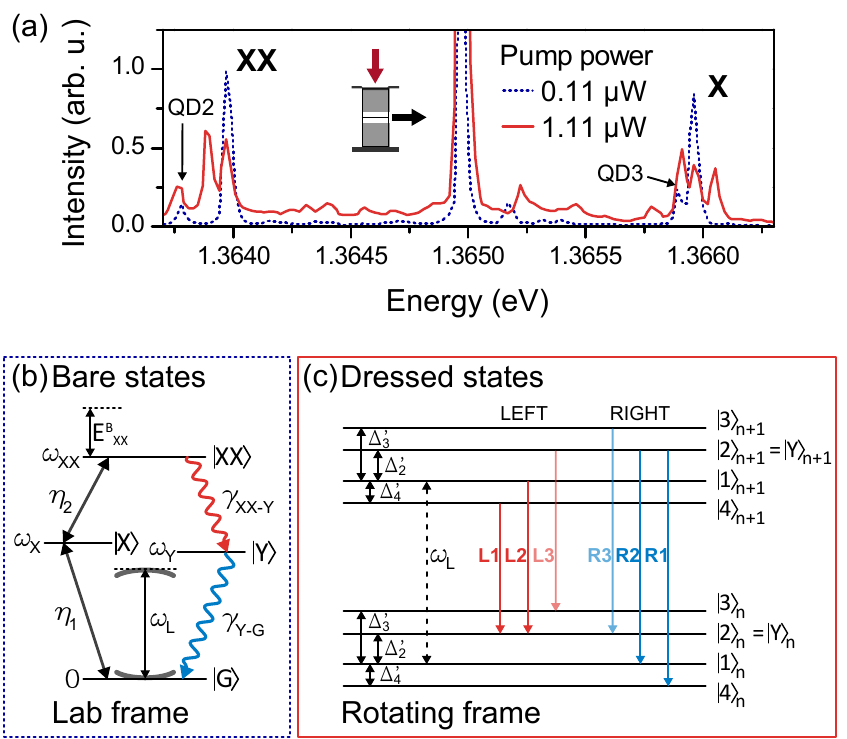}
        \caption{\label{Fig2}
      (Color online) (a) $y$-polarized  spectra $S_{\rm y}$ for two different pump intensities at a temperature of $T=6.8$\,K. Weak pump (blue, dotted): single biexciton XX and exciton X emission lines, with other emission lines probably come from additional QDs in the cavity. Strong pump (red, solid): dressed XX and X emission lines.
			(b) Energy level diagram of the bare QD states in the lab frame. The $x$-polarized transitions ($\left|XX\right>$-$\left|X\right>$-$\left|G\right>$) are coherently driven by a laser of frequency $\omega_L$ with amplitude $\eta_{1,2}$. The biexciton $\ket{XX}$ can decay via the $y$-polarized state $\ket{Y}$ by spontaneous emission with rates $\gamma_{XX-Y,Y-G}$ giving rise to the $y$-polarized spectra $S_{\rm y}$.
       (c) Dressed energy level diagram and allowed transitions in the rotating frame of the excitation laser $\omega_L$. The dressed states are labeled as $\left|1\right>$, $\left|2\right>$, $\left|3\right>$ and $\left|4\right>$. We have only plotted the 6 $y$-polarized transitions, which should be observable in our experiment (Fig.\,\ref{Fig1}d).}
\end{figure}

The sample used in the experiment is grown by molecular beam epitaxy and consists of self-assembled In$_{0.45}$Ga$_{0.55}$As QDs embedded in a high-Q micropillar resonator with a diameter of 2\,$\mu$m. The $\lambda$-cavity is composed of 26 and 30 pairs of distributed Bragg reflectors consisting of alternating $\lambda/4$ thick layers of GaAs and AlAs on the top and the bottom, respectively.
A thin aluminum mask on top of the micropillars reduces scattered laser light (Fig.~\ref{Fig1}(a)).
The sample is mounted inside a helium flow cryostat accessible via two microscope objectives ($\mbox{NA}=0.45$) in an orthogonal configuration, which allows for excitation and detection from both the top and the side of the micropillar.
A pair of crossed polarizers in front of the objectives further suppresses laser straylight.


To investigate basic optical parameters of the QD-cavity system we perform micro-PL spectroscopy, with the  pump laser at a frequency above the GaAs band edge. Using the top detection and two orthogonal positions of the polarizer, i.e., x- and y-polarization, we observe two cross-polarized fundamental modes (FMs) with an energy splitting of 320\,$\mu$eV and quality factors around 18500 (FM$_x$) and 10300 (FM$_y$), respectively (Fig.~\ref{Fig1}(b)).
This splitting might be caused by some ellipticity of the micropillars or anisotropic strain induced by remaining Benzocyclobutene (BCB) from the sample process.
Using the side detection, the QD emission becomes more distinctive, as the FM emission is strongly directional to the top but the QD emission is leaky to the side (Fig.~\ref{Fig1}(c)). Among the rest, we observe two emission lines at 1.364\,eV and 1.366\,eV, attributed to a biexciton-exciton cascade by means of second-order cross-correlation measurements---see supplementary material (SM).
At a temperature of 6.8\,K, these two lines are spectrally symmetric to the x-polarized FM$_x$ at 1.365\,eV enabling the cavity-enhanced coherent preparation of the biexciton state $\ket{XX}$ via two-photon absorption \cite{Brunner.Abstreiter.ea_1994,Flissikowski.Betke.ea_2004,Stufler.Machnikowski.ea_2006,Jayakumar.Predojevic.ea_2013,Muller.Bounouar.ea_2014}.
In the following experiment, we couple the resonant laser at a frequency $\omega_L\approx\omega^x_c$ from the top into the x-polarized cavity mode and detect the QD emission from the side. The $x$ and $y$ polarized excitons are almost aligned with the $x$ and $y$ polarized cavity states, respectively, and the resonant laser drives the $x$-polarized transitions (Fig.~\ref{Fig1}(d)). An orthogonal collection geometry only records the $y$-polarized emission.

Figure~\ref{Fig2}(a) shows two $y$-polarized spectra, one for moderate pump power (blue, dotted) and one for strong pump power (red, solid). At moderate pump power, the resonant excitation of the biexciton state results in the typical biexciton-exciton-cascade with almost equally intensities for both emission lines. The bright emission at 1.365\,eV corresponds to scattered laser light. Several other weaker emission lines are observable and might come from other QDs which are incoherently excited (typically through phonon-assisted processes). We labeled two lines with QD2 and QD3 to avoid confusion with the actually investigated QD. For a tenfold increased excitation power a distinct symmetric splitting of both transitions ($\left|XX\right>$-$\left|Y\right>$, $\left|Y\right>$-$\left|G\right>$) is observable. The magnitude of the splitting is $\Delta_{\Omega}=80\,\mu$eV ($\nu=19\,$GHz). In contrast, no other lines exhibit such a splitting, instead most of them are shifted away from the laser emission energy due to an AC-Stark effect.


To develop a simple physical understanding of the spectrum, we first model the transitions of a QD using a driven 4-level atom model shown in Fig.~\ref{Fig2}(b). The $x$-polarized transitions $\left|XX\right>$-$\left|X\right>$ and $\left|X\right>$-$\left|G\right>$ are coherently driven using a cw laser of frequency $\omega_L$ which couples to the transitions with Rabi frequency $\eta_2$ and $\eta_1$, respectively. The Hamiltonian of the system in a frame rotating at the laser frequency $\omega_L$ is given by
\begin{align}
\label{eq1a}
    &H\!= \hbar \Delta_4\ket{XX}\bra{XX} + \hbar \Delta_3\ket{X}\bra{X} + \hbar \Delta_{2}\ket{Y}\bra{Y} \nonumber\\
		&+ \hbar \eta_{2}(\smdag_{XX\mbox{-}X} +  \sm_{XX\mbox{-}X})  + \hbar \eta_{1}(\smdag_{X\mbox{-}G} +\sm_{X\mbox{-}G}),
\end{align}
where $\sm_{A\mbox{-}B}$ is the QD lowering operator from state $\ket{A}$ to $\ket{B},$ and $\Delta_4 = \omega_{XX}-2\omega_L$ and $\Delta_{2,3} = \omega_{Y,X}-\omega_L$ are the detunings for the biexciton and $x$ and $y$-polarized exciton states of energies $\omega_{XX}$, $\omega_X$ and $\omega_Y$, respectively, from the drive laser. The $X$ and $Y$ excitons are detuned by a small splitting $\delta_{XY}$~\cite{Muller.Fang.ea_2008}, due to the anisotropic exchange interaction. When the two-photon resonance condition ($\Delta_4 = 0$) is satisfied, $H_{\rm atom}$ can be  diagonalized to generate four new eigenstates $\ket{1, 2, 3, 4}$ with corresponding eigenvalues $\Delta'_{1,2,3,4}$. Since the $Y$ exciton, $\ket{Y}$, is not coupled to the drive, it is undressed and represents one of the eigenstates $\ket{2} = \ket{Y}$ with $\Delta'_2 = \Delta_2$. The other states $\ket{G}$, $\ket{X}$ and $\ket{XX}$ are simultaneously dressed by the drive to produce 3 new states $\ket{1,3,4}$ with eigenvalues $\Delta'_1 =0$ and $\Delta'_{3,4} = \abs{\frac{\Delta_3 \pm \sqrt{\Delta^2_3 + 4(\eta^2_1+\eta^2_2)}}{2}}$.
For a strong drive, multiple manifolds of these new states ($\ket{1,2,3,4}$) are formed (Fig.~\ref{Fig2}(c)) separated by drive frequency $\omega_L$. These different pump manifolds are identified by a subscript $n$, which represents the number of $x$-polarized photons from the coherent pump. Allowed transitions between the pump manifolds determine the total polarization unresolved spectra (see SM). In the experiment, only the $y$-polarized spectrum is measured and the transitions generating the spectra involve the $y$-polarized state ($\ket{2}$) (Fig.~\ref{Fig2}(c)). Hence, we now have 6 spectrally distinct, symmetric transitions about the drive frequency, marked `left' and `right'. These are the up and down transitions between state $\ket{Y}$ and the dressed states $\ket{1}$, $\ket{3}$ and $\ket{4}$. Note that the transition between state $\ket{Y}$ of 2 consecutive manifolds is polarization forbidden.


\begin{figure}[t!]
        \centering
                \includegraphics[width=0.48\textwidth]{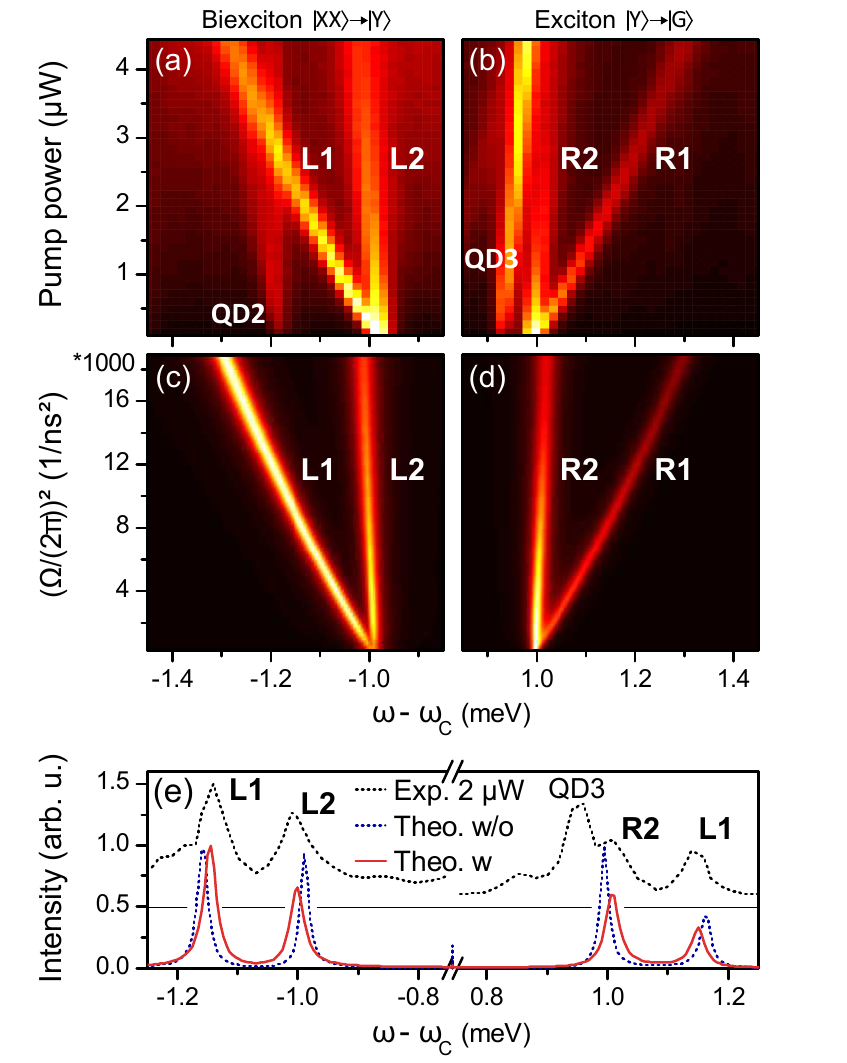}
                \caption{\label{Fig3}
                        (Color online) Color-scale maps of $S_{\rm y}$ showing dressed transitions around the original biexciton (a) and exciton (b) transitions, for increasing incident pump power.
                        (c) and (d) display the corresponding calculated normalized spectra. Simulation parameters (see text): $\gamma_{X\mbox{-}G, Y\mbox{-}G}$ = 0.56\,$\mu$eV, $\gamma_{XX\mbox{-}X,XX\mbox{-}Y}$ = 0.88\,$\mu$eV, $\gamma'_{XX\mbox{-}X,X\mbox{-}G,XX\mbox{-}Y,Y\mbox{-}G}$ = 8.2\,$\mu$eV, $\kappa_{x}$ = 74\,$\mu$eV and $\kappa_{y}$ = 132\,$\mu$eV, $\omega^x_c-\omega^y_c$ = 320\,$\mu$eV, $g_{X,Y}$= 26.7\,$\mu$eV and $g_2/g_1 = \sqrt{\gamma_{XX\mbox{-}X}/\gamma_{X\mbox{-}G}}$. The original exciton-ground transition ($\left|Y\right>$-$\left|G\right>$) is located +0.99\,meV and the biexciton-exciton transition ($\left|XX\right>$-$\left|Y\right>$) is located -0.99\,meV with respect to cavity at $\omega_c$ and $\delta_{XY}$= 25\,$\mu$eV. (e) Comparison of experimental spectrum for a pump power of 2\,$\mu$W (black, dotted) and the theoretical calculations with (red, solid) and without (blue, dotted) acoustic phonon contributions (see SM).
}
\end{figure}

Figures~\ref{Fig3}(a) and (b) display color-scale maps of the PL intensity for y-polarized biexciton and exciton transitions, respectively, as a function of emission energy and incident laser power P of the driving laser. The symmetric splitting $\Delta_{\Omega}$ of the two emission lines shows a linear dependence on the pump power and reaches values of up to $300\,\mu$eV. The labeling of the spectral lines corresponds to the transitions between the dressed states of the dressed 4-level atom as shown in Fig.~\ref{Fig2}(c) and the splitting $\Delta_{\Omega}$ can be estimated: The frequency of the two right Y transitions (R$_1$ and R$_2$) are $\omega_L+\Delta_2$ and 
$\omega_L+\Delta_2+\Delta'_4$, respectively, and their splitting $\Delta_{\Omega} =\Delta'_4 \approx \frac{\eta^2_1+\eta^2_2}{\Delta_3}$ grows linearly with power, at low powers. 


Figures~\ref{Fig3}(c) and (d) show the full theoretical calculation of y-polarized spectra $S_{\rm y}$ (see SM) including the effects of acoustic phonons and cavity interaction. The full Hamiltonian in the rotating frame of the laser is
\begin{align}
&H =  \hbar \Delta_4 \left| XX \rangle \langle XX \right| + \hbar \Delta_3 \left| X \rangle \langle X \right| + \hbar \Delta_2 \left| Y \rangle \langle Y \right|  \nonumber\\ 
&\!+ \sum_{P=x,y}\!\hbar \Delta_{cL}^P a^{\dagger}_P a_P  + \sum_q \hbar\omega_q b_q^{\dagger}b_q + \hbar \Omega (\adag_{x}  + \a_{x}) \nonumber\\
&\!+ \hbar g_{2}^x(\smdag_{XX\mbox{-}X} a_{x} + \adag_{x} \sm_{XX\mbox{-}X}) + \hbar g_{1}^x(\smdag_{X\mbox{-}G} a_{x} + \adag_{x} \sm_{X\mbox{-}G}) \nonumber \\
&\!+ \hbar g_{2}^y(\smdag_{XX\mbox{-}Y} a_{y} + \adag_{y} \sm_{XX\mbox{-}Y}) + \hbar g_{1}^y(\smdag_{Y\mbox{-}G} a_{y} + \adag_{y} \sm_{Y\mbox{-}G}) \nonumber \\  
&+ \sum_{S=X,Y,XX}\!\left| S \rangle \langle S \right| \sum_q\hbar \lambda_q^{S} \left(b^{\dag}_q + b_q\right),
\label{eq2}
\end{align}
where $\Delta^{x,y}_{cL} = \omega^{x,y}_c-\omega_L$ are the detunings for the $x,y$ polarized cavity of frequency $\omega^{x,y}_c$, described by lowering operators $\a_{x,y}$. The four QD transitions are coupled to the two cavities with the coupling $g^{x,y}_{1,2}$, where the superscript describes photon polarization and the subscripts 2 and 1 denote biexciton-to-exciton and exciton-to-ground state transition, respectively. To be consistent with the experiments (Fig.~\ref{Fig1}(d)), the $x$-polarized cavity is driven by an $x$-polarized laser with amplitude $\Omega$ and frequency $\omega_L$, which in turn drives the $x$-polarized transitions of the QD (see Fig.~\ref{Fig2}(b)). Since acoustic phonons are known to play an important role in excited QD-cavity systems~\cite{Ulrich.Ates.ea_2011,Roy.Hughes_2011,Kim.Shen.ea_2014}, the last three terms account for the electron-phonon interactions with the exciton and biexciton state at the level of the independent Boson model \cite{Wilson-Rae.Imamouglu_2002}. The coupling constant $\lambda^{P}_q$ couple the QD state $\ket{P}$ with the $q^{th}$ acoustic phonon mode (see SM). 
A useful approximation can be made to simplify $H$ (Eq.~\ref{eq2}), when the strongly driven $x$-polarized cavity has a large detuning ($\approx$ 1\,meV) from the coupled exciton ($\left|X\right>$-$\left|G\right>$) and biexciton ($\left|XX\right>$-$\left|X\right>$) transitions: an adiabatic approximation is made to describe the state of cavity with a complex number $\alpha$, where $\a_x\approx\alpha = \Omega/\left(i\Delta^x_{cL}+\frac{\kappa_x}{2}\right)$ (see SM), with $\kappa_x$ the $X$-cavity damping rate. This adiabatic elimination introduces effective Rabi pump fields $\eta_2$ and $\eta_1$, respectively, where $\eta_{1,2}= g^x_{1,2}\alpha$, similar to Eq.~\ref{eq1a}. The $X$-cavity thus essentially filters the original cw cavity drive $\Omega$. This approximation allows us to determine the $X$-cavity photon number analytically, e.g., for a particular line splitting $\Delta_{\Omega}$, from $N_c =\braket{\adag \a}_X\approx \abs{\alpha}^2 = \frac{\Delta_{\Omega}(\Delta_{\Omega}+\Delta_3)}{(g^x_2)^2+(g^x_1)^2}$ (see SM). Thus for the maximum splitting of $\hbar\Delta_{\Omega} = 300 \mu$eV in Fig.~\ref{Fig3}, $N_c\approx$ 212. Note that while the $y$-polarized cavity is not directly driven, it still strongly assists the radiative decay of the QD using $y$-polarized transitions, especially in presence of phonons. A polaron transformed master equation  for the QD-cavity system is then derived using $H$, which includes phonon interactions nonperturbatively. Additional incoherent processes like radiative decay ($\gamma$) and pure dephasing ($\gamma'$) of the QD transitions and cavity damping ($\kappa_y$) are also included  using standard Lindblad decay terms, $L(O) = \frac{\Gamma}{2}(2O\rho O^{\dagger} - O^{\dagger} O\rho -  \rho O^{\dagger} O)$, with $\Gamma$ the  rate. 


\begin{figure}
        \centering
                \includegraphics[width=0.99\columnwidth]{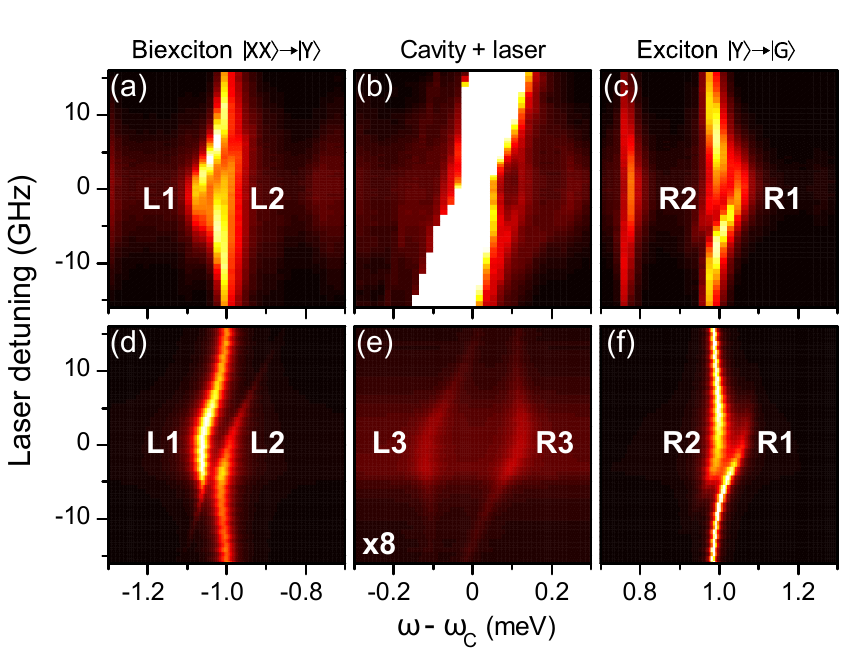}
                \caption{\label{Fig4}
                        (Color online) Color-scale map of the PL spectra as a function of energy and laser detuning around the original biexciton (a), cavity (b) and exciton (c) emission lines at a temperature of $20.2$\,K. When tuning the pump laser through the cavity we observe an anticrossing for both emission lines X and XX. The lines are labeled corresponding to the transitions in Fig.~\ref{Fig1}(b).
                        (d) - (f) display the calculated normalized emission spectra. Parameters are the same as in Fig.~\ref{Fig3} and incorporate the frequency shifts due to $T=20.2\,$K (see text).
                }
\end{figure}

Phonon interactions and incoherent losses determine the linewidth and the oscillator strength of the different spectral transitions. Phonons play an important role in determining the final intensities in the spectrum as can be observed by comparing an experimental measurement (black dotted line at a pump power of $2\,\mu$W) against theoretical calculations with (red solid) and without phonons (blue dashed) in Fig.~\ref{Fig3}(e).
Without phonons the oscillator strength of the transitions to the right of the laser (cavity) are relatively stronger (i.e. $R2 \approx L2, L1$).
The corrected oscillator strength with phonons can be qualitatively explained by understanding the asymmetric phonon-dressed Mollow spectra from single excitons;  due to phonon-induced excitation processes, transitions to the left of the pump laser are always fed more strongly than transitions to the right. This is because at low temperatures, phonon emission is more probable than absorption. Coupling to the $Y$-cavity strengthens the left transitions further, since it allows faster phonon-mediated decay of the coupled right transitions. 


Figures~\ref{Fig4}(a)-(c) present $S_{\rm y}$ measurements around the original biexciton, cavity, and exciton transitions, respectively, when the frequency of the driving laser is scanned through the cavity resonance at fixed pump power of 6.6\,$\mu$W.
The experiment was performed at a temperature of roughly 20.2\,K to spectrally tune the quantum dot emission and to re-establish the two-photon resonance condition \footnote{During several cooling and heating cycles some strain or field changes in the micropillar caused a blue-shift of the QD emission with respect to the cavity mode. Therefore, the following experiments were performed at a temperature of 20.2\,K to tune the QD emission and to almost re-establish the two-photon resonance condition. Hence, the emission lines are slightly shifted to lower energies compared to the previously presented measurements.}.
Figures~\ref{Fig4}(d)-(f) show the respective normalized spectra calculations. By tuning the laser frequency through the cavity we observe an anticrossing like behavior for both the exciton and the biexciton emission lines at the same time. The splitting reaches values of up to $\Delta_{\Omega}=60\,\mu$eV at zero detuning. The on-resonance ($\omega_L=\omega^x_c$) doublets (Fig.~\ref{Fig4} (a, c)) correspond to distinct dressed transitions in Fig.~\ref{Fig3} (a, b).
The cavity not only aids field enhancement but also acts as spectral filter for the pump laser. Thus, the effective drive amplitude depends on the detuning, i.e., $\eta_{1,2} =  g_{1,2}\Omega/\left(i\Delta^x_{cL}+\frac{\kappa_x}{2}\right)$. This becomes noticeable in the re-shaped anticrossing.
In Fig.~\ref{Fig4}(b) we observe several weaker lines located around the cavity frequency. These lines follow the expected trend of the corresponding calculated transitions $L_3, R_3$ in Fig.~\ref{Fig4}(e), but we cannot exclude contributions from other QD emission lines. The strong scattered laser light prohibits to clarify this with reasonable certainty.


In conclusion, we have demonstrated a unique dressed biexciton-exciton system by cavity-enhanced two-photon cw excitation of a semiconductor QD. The dressed states are studied by $\mu$PL  spectroscopy and reveal a characteristic emission pattern of four distinct emission lines. The splitting of the biexciton-exciton ($\left|XX\right>$-$\left|Y\right>$) and exciton-ground state ($\left|Y\right>$-$\left|G\right>$) transition show linear dependence on the applied power of the driving laser field.
We introduced a dressed four-level system to describe and calculate the measurements theoretically, and we have also highlighted the importance of electron-phonon scattering to the model to match more quantitatively using a polaron master equation approach.

\begin{acknowledgments}
The authors would like to thank M. Emmerling for expert sample fabrication.
We acknowledge financial support of the Deutsche Forschungsgemeinschaft (DFG) within the SFB/TRR21 and the projects MI 500/23-1 and Ka2318/4-1.
and the Natural Sciences and Engineering Research Council of Canada.
\end{acknowledgments}


\bibliography{CATPE}

\printindex

\end{document}